\documentclass[aps,pra,twocolumn,reprint,superscriptaddress,nofootinbib]{revtex4-2}

\usepackage[svgnames,table,xcdraw]{xcolor}
\usepackage{amsmath}
\usepackage{microtype}
\usepackage{amssymb}
\usepackage{amsmath}
\usepackage{braket}
\usepackage{graphicx}
\usepackage{booktabs}
\usepackage{bm}
\usepackage{dsfont}
\usepackage{hyperref}
\usepackage[caption=false]{subfig}
\usepackage{amsthm}

\usepackage{xfrac}
\usepackage{enumitem}
\usepackage{array}

\hypersetup{colorlinks,linkcolor={DarkBlue},citecolor={DarkBlue},urlcolor={DarkBlue}}

\let\originalleft\left
\let\originalright\right
\renewcommand{\left}{\mathopen{}\mathclose\bgroup\originalleft}
\renewcommand{\right}{\aftergroup\egroup\originalright}

\usepackage{nag}
\usepackage{graphicx}
\usepackage{amsthm}
\usepackage{qcircuit}
\usepackage{mathtools}
\usepackage{bm}
\usepackage[ruled,vlined]{algorithm2e}
\usepackage{braket}
\usepackage{datetime}
\usepackage{stackrel}
\usepackage{dsfont}
\usepackage{qcircuit}
\usepackage[english]{babel}
\usepackage{units}

\usepackage{tikz}
\usetikzlibrary{backgrounds,decorations.pathreplacing}

\usepackage{chngcntr}
\counterwithout{equation}{section}

\usepackage[normalem]{ulem}
\usepackage{slashed}
\usepackage{soul}

\usepackage{xcolor}
\usepackage{amssymb}
\usepackage{amsmath}
\usepackage{amsthm}
\usepackage{braket}
\usepackage{mathdots}

\usepackage{makeidx}
\makeindex

\usepackage{soul}
\usepackage{xargs}
\newcommandx{\cmnote}[2][1=]{\linespread{1.0}\todo[linecolor=red,backgroundcolor=red!25,bordercolor=red,#1]{#2}}

\let\underline\ul

\allowdisplaybreaks[4]

 \index{} \index{} \index{} \index{} \index{} \index{} \index{} \index{}%%
\makeatletter

\newcommand{\ringplus}{\mathbin{\text{\@ringplus}}}

\newcommand{\@ringplus}{%
  \ooalign{\hidewidth\raise1.3ex\hbox{\tiny$\circ$}\hidewidth\cr$\m@th+$\cr}%
}

\newcommand{\ringminus}{\mathbin{\text{\@ringminus}}}

\newcommand{\@ringminus}{%
  \ooalign{\hidewidth\raise0.9ex\hbox{\tiny$\circ$}\hidewidth\cr$\m@th-$\cr}%
}
\makeatother
 \index{} \index{} \index{} \index{} \index{} \index{} \index{} \index{}%%

\newcommand{\logical}[0]{L}

\DeclareFontFamily{U}{wncy}{}
\DeclareFontShape{U}{wncy}{m}{n}{<->wncyr10}{}
\DeclareSymbolFont{mcy}{U}{wncy}{m}{n}
\DeclareMathSymbol{\Sh}{\mathord}{mcy}{"58}

 \index{} \index{} \index{} \index{} \index{} \index{} \index{} \index{}%%
 \index{} \index{} \index{} \index{} \index{} \index{} \index{} \index{}%%

 \index{} \index{} \index{} \index{} \index{} \index{} \index{} \index{}%%
 \index{} \index{} \index{} \index{} \index{} \index{} \index{} \index{}%%
\usepackage{graphicx}
\usepackage{graphicx}
\usepackage{multirow}
\usepackage{hhline}
\xyoption{color}
\xyoption{line}

\newcommandx*\bsbal[3][1=black, 3=->]{\ar @[#1]@{#3} [#2,0] \qw}

\newcommandx*\varbs[5][1=black, 3=\theta,4=0.5,5=->]{\ar @[#1]@{#5}^(#4){#3} [#2,0] \qw}

\newcommandx*\lblline[3][3=0.5]{\ar @{-}^(#3){#1} [#2,0]}
\newcommandx*\ctrlg[3][3=0.5]{ \raisebox{-3pt}{$\bullet$}  \ar @{-}^(#3){#1} [#2,0] \qw }
\newcommandx*\ctrlog[2]{\controlo \ar @{-}^{#1} [#2,0] \qw}
\newcommandx*\ctrlodash[1]{\controlo \ar @{-} [#1,0] \ar @[black]@{.} [0,-1]}

\begin{document}

\title{%
  \texorpdfstring
  {LiDMaS: Architecture-Level Modeling of Fault-Tolerant \\Magic-State Injection in GKP Photonic Qubits}
  {LiDMaS: Architecture-Level Modeling of Fault-Tolerant Magic-State Injection in GKP Photonic Qubits}
}
\def \affGatech {College of Computing, Georgia Institute of Technology, Atlanta, GA 30332 USA}
\def \schrosim {Independent Quantum Architecture Researcher \& Software Developer, SchroSIM Quantum Software Project}
\author{Dennis Delali Kwesi Wayo}
\affiliation{\affGatech}
\affiliation{\schrosim}
\email{dwayo3@gatech.edu}
\date{\today}

\begin{abstract}
Fault-tolerant quantum computation in photonic architectures requires efficient preparation of high-fidelity logical magic states under realistic constraints of finite squeezing and photon loss. We present LiDMaS (Lightweight Density-Matrix Simulator), an architecture-level study of logical $T$-gate magic-state preparation in Gottesman--Kitaev--Preskill (GKP)-encoded photonic qubits using a repeat-until-success (RUS) injection protocol combined with outer surface-code protection. A density-matrix simulator based on standard numerical linear algebra is employed, mapping finite squeezing to effective logical dephasing, incorporating logical depolarizing noise, and treating photon loss as a heralded erasure process. Parameter sweeps are performed over squeezing values from $8$ to $16$~dB, baseline loss probabilities between $0.005$ and $0.03$, and surface-code distances $d=1,3,5,$ and $7$. Across this regime, RUS success probabilities range from about $0.90$ to $0.99$, with average injection overhead between $1.15$ and $1.21$ rounds per successful attempt. After outer-code protection, logical fidelities reach $F_{\mathrm{log}} \approx 0.765\text{--}0.796$, exhibiting weak sensitivity to moderate photon loss but strong monotonic dependence on squeezing. Sensitivity analysis identifies finite squeezing as the dominant continuous error source, while loss primarily impacts heralded failure rates. Phase-boundary diagrams determine minimum squeezing requirements to achieve success probability $\ge 0.95$ and logical fidelity $\ge 0.79$ as a function of code distance, providing quantitative design guidance for scalable photonic fault-tolerant architectures.
\end{abstract}

\maketitle

\section{Introduction}

Photonic quantum computing has emerged as a leading platform for scalable fault-tolerant quantum information processing due to its intrinsic advantages, including low decoherence, room-temperature operation, and the natural compatibility of photons with long-distance communication \cite{rudolph2017optimistic,slussarenko2019photonic}. However, achieving universal, fault-tolerant quantum computation in photonic systems requires overcoming fundamental challenges associated with non-Clifford gate implementation, photon loss, and the finite quality of experimentally realizable quantum states \cite{nickerson2014freely,bourassa2021blueprint}. Among the most promising approaches to addressing these challenges is the use of the Gottesman--Kitaev--Preskill (GKP) encoding, which enables the embedding of qubit degrees of freedom into continuous-variable (CV) photonic modes and allows Gaussian operations to correct small displacement errors \cite{gottesman2001encoding,menicucci2014fault}.

Early theoretical work by Menicucci \emph{et al.}~\cite{menicucci2014fault} established rigorous fault-tolerance thresholds for measurement-based quantum computation using GKP-encoded cluster states under finite squeezing. That analysis demonstrated that, in principle, universal fault-tolerant quantum computation is achievable with sufficiently high squeezing and idealized Gaussian operations. Subsequent studies have refined this picture by incorporating more realistic noise models and decoding strategies, including the use of analog measurement information to improve error correction performance at moderate squeezing levels~\cite{fukui2018high,noh2020fault}. Collectively, these works have established GKP encoding as a central ingredient in photonic fault-tolerance roadmaps~\cite{bourassa2021blueprint}.

Despite this progress, a critical gap remains between decoder-level analyses of GKP error correction and the system-level requirements of universal fault-tolerant computation. In particular, non-Clifford gate synthesis, most notably the implementation of the logical $T$ gate, is indispensable for universality and typically relies on magic-state preparation~\cite{bravyi2005universal,howard2014contextuality}. In photonic architectures, magic-state generation is subject to the same physical constraints as logical qubit storage and Clifford operations, including finite squeezing, residual logical noise, and photon loss~\cite{nickerson2014freely,rudolph2017optimistic}. Yet, most existing analyses of GKP-based fault tolerance focus either on asymptotic thresholds or on error suppression under stabilizer operations, without explicitly modeling magic-state injection protocols, repeat-until-success (RUS) dynamics, or the interplay between heralded loss and logical fidelity~\cite{campbell2012magic}.

Magic-state preparation has been studied extensively in qubit-based architectures, particularly through distillation protocols and measurement-based constructions~\cite{campbell2009structure,bravyi2005universal,litinski2019game}. These works provide deep insights into overhead scaling and resource optimality, but they generally assume abstract noise models and do not account for photonic-specific effects such as squeezing-dependent dephasing or loss-induced erasure. In photonic systems, photon loss is both ubiquitous and, crucially, often heralded, allowing failed operations to be detected and discarded rather than corrupting logical states~\cite{ralph2005loss}. This distinguishing feature suggests that the dominant constraints on logical gate quality may differ substantially from those in matter-based qubit platforms, motivating a dedicated analysis tailored to photonic architectures.

At the architectural level, surface-code-based protection has been proposed as a powerful outer layer to suppress residual logical errors in photonic systems. Notably, Nickerson, Fitzsimons, and Benjamin~\cite{nickerson2014freely} demonstrated that heralded photon loss can be efficiently tolerated by erasure-aware decoding, significantly relaxing loss requirements for large-scale quantum computation. However, their analysis focused primarily on Clifford operations and logical error rates, leaving open the question of how non-Clifford gate synthesis, and in particular magic-state fidelity, behaves under realistic loss and noise conditions~\cite{fowler2012surface}. Understanding this behavior is essential for assessing the viability of photonic fault-tolerant architectures beyond threshold arguments.

Recent architectural reviews and roadmaps, such as those by Rudolph~\cite{rudolph2017optimistic} and subsequent system-level analyses~\cite{bourassa2021blueprint}, have emphasized squeezing as a key hardware bottleneck for photonic quantum computing, while acknowledging uncertainty in how squeezing, loss, and code distance jointly constrain system performance. While these studies provide valuable qualitative guidance, quantitative design rules linking experimentally tunable parameters to operational metrics such as success probability, overhead, and logical fidelity remain scarce~\cite{webster2022universal}. In particular, it is unclear to what extent photon loss versus finite squeezing limits the quality of successfully prepared logical magic states, and how outer-code protection reshapes these trade-offs.

In this work, these open questions are addressed via a systematic architecture-level investigation of logical $T$-gate magic-state preparation in GKP-encoded photonic qubits. We refer to this framework as LiDMaS (Lightweight Density-Matrix Simulator). Rather than performing explicit continuous-variable wavefunction simulations or decoder-level modeling, we adopt a deliberately simplified yet physically informed framework in which logical qubits are represented by $2\times2$ density matrices and noise processes are modeled as effective logical channels. Finite GKP squeezing is mapped to an effective Pauli-$Z$ dephasing channel, residual imperfections are captured by logical depolarizing noise, and photon loss is treated as a heralded erasure process that aborts failed attempts without degrading successful outcomes. This abstraction allows us to efficiently explore a wide parameter space while retaining direct correspondence with experimentally relevant quantities. The novelty is the explicit coupling of a RUS non-Clifford injection workflow to a photonic noise model with heralded loss and an outer-code abstraction, enabling direct estimates of success probability, overhead, and logical fidelity rather than only asymptotic thresholds.

Focus is placed on a repeat-until-success (RUS) magic-state injection protocol combined with an outer surface-code abstraction, enabling quantification of key operational metrics, including success probability within a finite round cap, average injection overhead, and logical magic-state fidelity after outer-code protection. By sweeping squeezing levels, baseline loss probabilities, and surface-code distances, sensitivity maps and phase-boundary diagrams are constructed to reveal how hardware quality and error correction jointly determine performance. Importantly, this approach disentangles the roles of squeezing and loss: while loss primarily affects heralded failure rates and overhead, finite squeezing emerges as the dominant factor limiting logical fidelity.

The main contributions of this work are threefold. First, a unified architecture-level model is provided that explicitly incorporates finite squeezing, depolarizing noise, and heralded photon loss into logical magic-state preparation, without relying on continuous-variable simulation frameworks or quantum software libraries such as PennyLane. Second, sensitivity analysis demonstrates that logical magic-state fidelity is remarkably robust to moderate loss once heralded erasure is accounted for, but remains strongly dependent on squeezing across all studied code distances. Third, quantitative phase-boundary diagrams are introduced to identify the minimum squeezing required to simultaneously achieve high success probability and high logical fidelity as a function of loss and code distance, thereby offering practical design guidance for photonic quantum computing platforms.

By bridging the gap between decoder-centric analyses and system-level gate synthesis, this study clarifies the dominant constraints on fault-tolerant photonic quantum computation and provides actionable insights for experimental and architectural development. The results underscore the central importance of squeezing improvements for enhancing logical gate quality, while highlighting the comparatively benign role of photon loss in heralded, RUS-based photonic architectures.

\section{Methods}
\label{sec:methods}

\subsection{Architecture-Level Modeling Framework}

The results presented in this work are obtained using a custom, lightweight density-matrix simulation framework designed to capture the dominant logical error mechanisms relevant to fault-tolerant photonic quantum architectures~\cite{bourassa2021blueprint,divincenzo2009fault}. Rather than performing explicit continuous-variable wavefunction simulations of Gottesman–Kitaev–Preskill (GKP) states or relying on quantum software frameworks such as PennyLane~\cite{bergholm2018pennylane}, an architecture-level approach is adopted in which encoded qubits are represented by $2\times2$ density matrices and logical operations are modeled using standard matrix algebra.This deliberate abstraction enables efficient numerical exploration of design trade-offs while preserving the essential structure of logical noise propagation and error correction~\cite{noh2020fault}.

\begin{figure*}[t]
    \centering
    \includegraphics[width=0.6\linewidth]{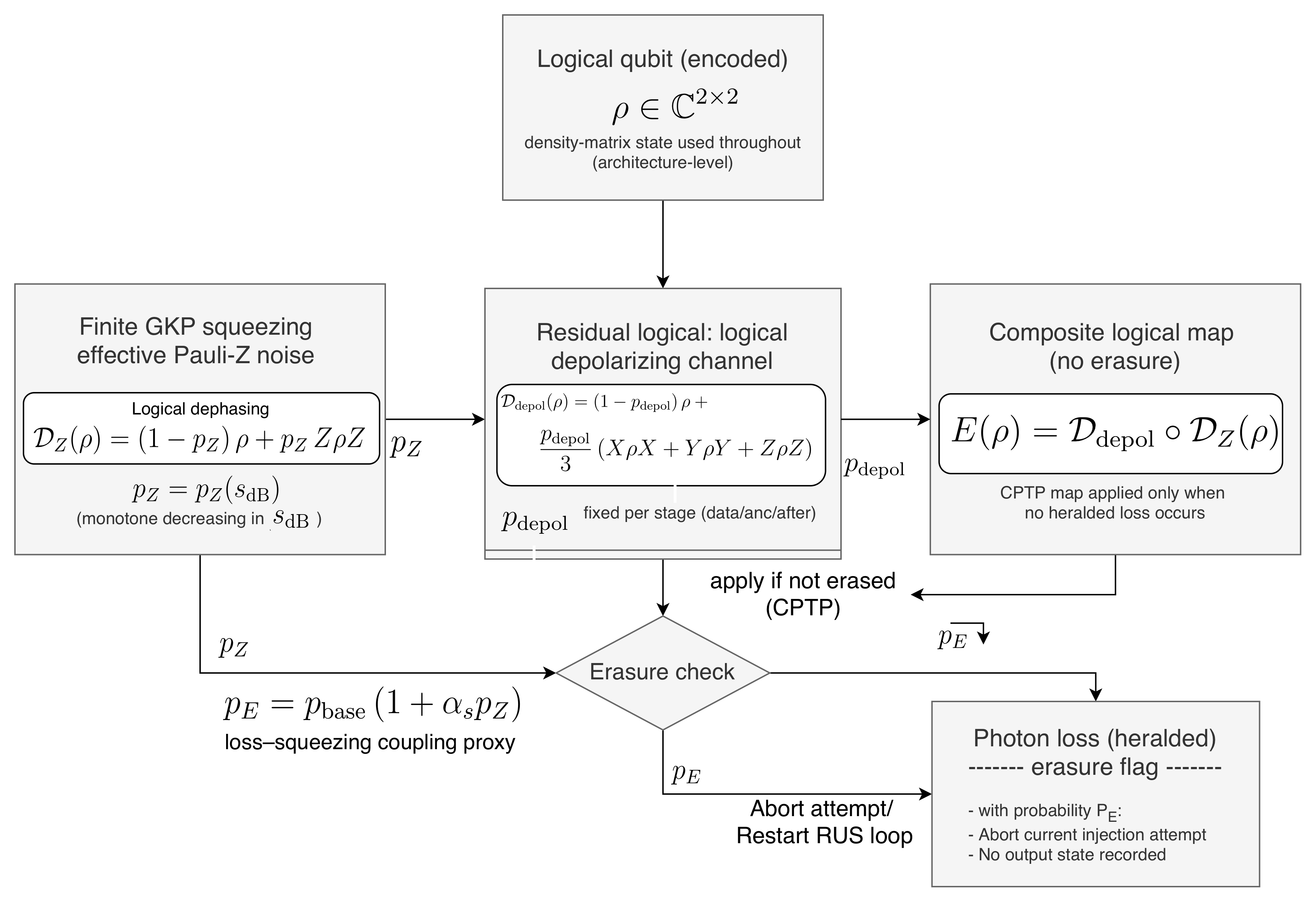}
    \caption{
    Architecture-level logical noise model used throughout this work. Encoded qubits are represented as $2\times2$ density matrices and evolve under effective logical noise channels.
    Finite GKP squeezing induces Pauli-$Z$ dephasing with rate $p_Z(s)$, while residual imperfections are captured by a logical depolarizing channel with rate $p_{\mathrm{depol}}$. Photon loss is treated as a heralded erasure process with probability $p_E$, aborting the current repeat-until-success injection attempt. When no erasure occurs, the composite completely positive trace-preserving map $\mathcal{E}(\rho)$ is applied. This abstraction avoids explicit continuous-variable simulation while preserving the dominant fault-tolerant error mechanisms. }
    \label{fig:logical_noise_model}
\end{figure*}

Finite squeezing of approximate GKP states is incorporated through an effective logical dephasing channel, parameterized by a squeezing-dependent error probability calibrated to reproduce the expected monotonic improvement with increasing squeezing~\cite{fukui2018high,hastrup2021improved}. Additional logical depolarizing noise is included to model residual non-Pauli errors arising from imperfect Clifford operations and finite stabilizer-cycle fidelity~\cite{gottesman1997stabilizer}. Photon loss, the dominant hardware impairment in photonic platforms, is treated as a heralded erasure process that terminates the current magic-state injection attempt without contributing unheralded logical errors~\cite{nickerson2014freely}. This modeling choice reflects the strong loss-detectability inherent to photonic architectures and allows the separation of failure probability from logical-state quality. In implementation, all noise probabilities are clipped to the physical interval $[0,1]$, and the squeezing-to-dephasing proxy is further capped at $0.5$ to preserve complete positivity. The resulting logical noise model is summarized in Fig.~\ref{fig:logical_noise_model}.

At the logical level, non-Clifford gate implementation is realized using a repeat-until-success (RUS) $T$-gate magic-state injection protocol~\cite{paetznick2013repeat}. The output of the injection stage is subsequently protected by an outer quantum error-correcting code, modeled here using a surface-code-inspired logical error scaling law~\cite{fowler2012surface}. This layered abstraction, logical GKP encoding, RUS-based magic-state preparation, and outer-code suppression of residual Pauli errors, provides a transparent and computationally efficient framework for evaluating fault-tolerant performance across wide ranges of squeezing, loss, and code distance~\cite{bourassa2021blueprint}.

\subsection{Logical noise model}

Each logical qubit is described by a density matrix $\rho \in \mathbb{C}^{2 \times 2}$ and is subject to three independent effective noise channels:
\begin{enumerate}
    \item phase noise induced by finite GKP squeezing,
    \item depolarizing noise capturing residual Clifford and stabilizer imperfections,
    \item photon-loss-induced erasure.
\end{enumerate}

These effects are modeled using the composite map in Equation~\ref{eq:1}.
\begin{equation}
\mathcal{E}(\rho) =
\begin{cases}
\text{erasure},
& \text{with probability } p_{\mathrm{E}}, \\[4pt]
\mathcal{D}_{\mathrm{depol}} \circ \mathcal{D}_{Z}(\rho),
& \text{otherwise}.
\end{cases}
\label{eq:1}
\end{equation}
where $p_{\mathrm{E}}$ is the erasure probability~\cite{gottesman1997stabilizer,nickerson2014freely,noh2020fault}.

\subsubsection{Finite-squeezing-induced dephasing}

Finite-energy GKP states exhibit residual phase uncertainty that manifests as effective logical dephasing after GKP error correction. This effect is modeled as a Pauli-Z channel (Equation~\ref{eq:2}),
\begin{equation}
\mathcal{D}_{Z}(\rho) = (1 - p_Z)\rho + p_Z Z \rho Z,
\label{eq:2}
\end{equation}
where $p_Z$ is an effective dephasing probability.

To relate this to the experimental squeezing parameter $s$ (expressed in dB), a monotonic proxy mapping is introduced in Equation~\ref{eq:3}.
\begin{equation}
p_Z(s) = \min\!\left(0.5,\, \alpha_s e^{-\beta s}\right),
\label{eq:3}
\end{equation}
where $\alpha_s$ and $\beta$ are phenomenological constants. This form captures the exponential suppression of GKP logical errors with increasing squeezing and is consistent with analytic and numerical studies of finite-energy GKP codes.

\subsubsection{Depolarizing noise}

Residual logical noise not directly associated with squeezing is modeled as a depolarizing channel (Equation \ref{eq:4}),
\begin{equation}
\mathcal{D}_{\mathrm{depol}}(\rho)
= (1 - p_{\mathrm{dep}})\rho
+ \frac{p_{\mathrm{dep}}}{3}
\left(
X\rho X + Y\rho Y + Z\rho Z
\right).
\label{eq:4}
\end{equation}
where $p_{\mathrm{dep}}$ is fixed independently for data qubits, ancilla qubits, and post-injection outputs.

This term captures logical errors arising from imperfect Clifford gates, syndrome extraction, and measurement errors not explicitly resolved in the model.

\subsubsection{Photon loss and erasure}

Photon loss in photonic architectures is modeled as a heralded erasure process. When an erasure event occurs, the current magic-state injection attempt is aborted and flagged as failed. Importantly, erasures do not propagate logical Pauli errors into successful outcomes.

The baseline erasure probability $p_{\mathrm{base}}$ is coupled to finite squeezing via Equation \ref{eq:5}
\begin{equation}
p_{\mathrm{E}} = p_{\mathrm{base}} \left( 1 + \alpha_{\mathrm{LS}} p_Z \right),
\label{eq:5}
\end{equation}
where $\alpha_{\mathrm{LS}}$ quantifies loss–squeezing coupling. This reflects the increased sensitivity of low-squeezing GKP states to photon loss.

\subsection{Logical $T$-gate magic-state injection}

The standard magic-state injection protocol based on the $|A\rangle$ state in Equation~\ref{eq:6} is considered.
\begin{equation}
|A\rangle = T|+\rangle = \frac{1}{\sqrt{2}} \left( |0\rangle + e^{i\pi/4}|1\rangle \right),
\label{eq:6}
\end{equation}
where $T = \mathrm{diag}(1, e^{i\pi/4})$.

\subsubsection{Injection circuit}

\begin{figure*}[t]
    \centering
    \includegraphics[width=0.6\linewidth]{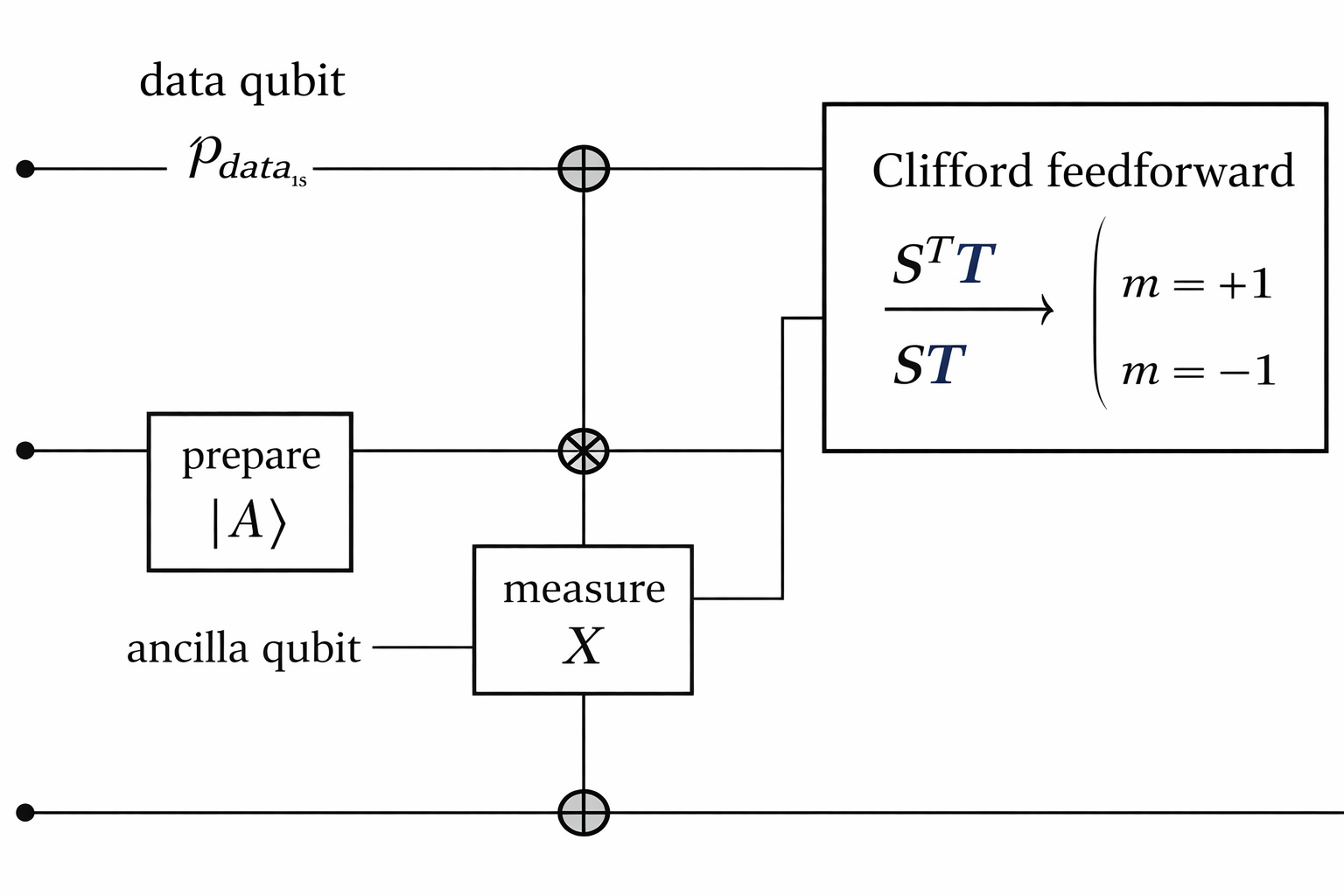}
    \caption{
    Schematic of the logical $T$-gate magic-state injection protocol used in this work. A data qubit encoded at the logical level interacts with an ancilla prepared in the magic state $|A\rangle = T|+\rangle$ via a CNOT gate, followed by an $X$-basis measurement of the ancilla. Conditioned on the measurement outcome, a Clifford feedforward operation ($S$ or $S^\dagger$) is applied to the data qubit. One measurement branch implements the desired logical $T$ gate (up to Clifford equivalence), while the other branch requires repetition, yielding a repeat-until-success (RUS) protocol. Photon loss events are treated as heralded erasures that abort the current attempt. }
    \label{fig:rus_t_injection}
\end{figure*}

The injection circuit operates on a data qubit $\rho_{\mathrm{data}}$ and an ancilla prepared in $|A\rangle$, and consists of:
\begin{enumerate}
\item preparation of the ancilla state,
\item a CNOT gate (data as control, ancilla as target),
\item measurement of the ancilla in the $X$ basis,
\item Clifford feedforward on the data qubit (see Figure \ref{fig:rus_t_injection}).
\end{enumerate}

Conditioned on the measurement outcome $m \in {+1,-1}$, the data qubit undergoes Equation \ref{eq:7}
\begin{equation}
\rho \;\rightarrow\;
\begin{cases}
S^\dagger T \rho T^\dagger S,
& m = +1, \\[4pt]
S T^\dagger \rho T S^\dagger,
& m = -1.
\end{cases}
\label{eq:7}
\end{equation}

One measurement branch is designated as the successful $T$-gate realization up to Clifford equivalence, while the complementary branch is treated as unsuccessful and triggers repetition. This choice reflects a standard architecture-level abstraction used in RUS analyses.

\subsubsection{Repeat-until-success abstraction}

The injection protocol is embedded within a repeat-until-success (RUS) loop. Each attempt may result in:
\begin{itemize}
\item erasure (photon loss),
\item unsuccessful Clifford branch,
\item successful $T$-gate injection.
\end{itemize}

The number of attempts $R$ required follows a geometric distribution truncated at a maximum allowed number of rounds $R_{\max}$. Metrics reported include (Equation \ref{eq:8}):

\begin{subequations}
\begin{align}
P_{\mathrm{succ}}
&= \Pr(R \le R_{\max}), \\
\langle R \rangle_{\mathrm{succ}}
&= \mathbb{E}\!\left[ R \mid R \le R_{\max} \right], \\
F_{\mathrm{inj}}
&= \left\langle
F\!\left(\rho_{\mathrm{out}},\, T\ket{\psi}\right)
\right\rangle_{\mathrm{succ}} .
\end{align}
\label{eq:8}
\end{subequations}

\begin{algorithm}[t]
\caption{Architecture-level RUS $T$-state injection sweep}
\KwIn{Squeezing values $\{s\}$, loss bases $\{p_{\mathrm{base}}\}$, distances $\{d\}$, trials $N$, round cap $R_{\max}$}
\KwOut{Metrics: $P_{\mathrm{succ}}$, $\langle R\rangle_{\mathrm{succ}}$, $F_{\mathrm{inj}}$, $F_{\mathrm{log}}$}
\ForEach{$p_{\mathrm{base}} \in \{p_{\mathrm{base}}\}$}{
  \ForEach{$s \in \{s\}$}{
    $p_Z \leftarrow \min(0.5,\, \alpha_s e^{-\beta s})$\;
    $p_E^{\mathrm{data}} \leftarrow \mathrm{clip}\!\left(p_{\mathrm{base}}(1+\alpha_{\mathrm{LS}}p_Z)\right)$\;
    $p_E^{\mathrm{anc}} \leftarrow \mathrm{clip}\!\left(1.5\,p_{\mathrm{base}}(1+\alpha_{\mathrm{LS}}p_Z)\right)$\;
    \ForEach{$d \in \{d\}$}{
      Initialize counters and accumulators\;
      \For{$i=1$ \KwTo $N$}{
        \For{$r=1$ \KwTo $R_{\max}$}{
          Apply logical noise to data and ancilla; if erasure then abort trial\;
          Apply CNOT, measure ancilla in $X$ basis, apply Clifford feedforward\;
          Apply post-noise; if erasure then abort trial\;
          \If{success branch}{
            Compute $F_{\mathrm{inj}}$ against $T|\psi\rangle$\;
            $p_{\mathrm{phys}} \leftarrow w_Z p_Z + w_P p_{\mathrm{dep}}$\;
            $p_L \leftarrow A \left(\frac{p_{\mathrm{phys}}}{p_{\mathrm{th}}}\right)^{(d+1)/2}$, capped at $1/2$ if $p_{\mathrm{phys}}\ge p_{\mathrm{th}}$\;
            $F_{\mathrm{log}} \leftarrow (1-p_L)F_{\mathrm{inj}} + p_L/2$\;
            Record success and break\;
          }
        }
      }
      Compute $P_{\mathrm{succ}}$, $\langle R\rangle_{\mathrm{succ}}$, and averages\;
    }
  }
}
\end{algorithm}

\subsection{Outer-code abstraction: surface-code protection}

Successful injection produces a logical state that is further protected by an outer surface code of distance $d$. Rather than explicitly simulating stabilizer cycles and decoding, a standard scaling-law abstraction for the logical error rate is adopted (Equation~\ref{eq:9}):

\begin{equation}
p_L(p_{\mathrm{phys}}, d)
= A \left( \frac{p_{\mathrm{phys}}}{p_{\mathrm{th}}} \right)^{(d+1)/2},
\qquad p_{\mathrm{phys}} < p_{\mathrm{th}},
\label{eq:9}
\end{equation}

where $p_{\mathrm{th}}$ is the threshold error rate and $A$ is a prefactor of order unity. In implementation, if $p_{\mathrm{phys}} \ge p_{\mathrm{th}}$ the logical failure probability is pessimistically capped at $p_L=1/2$, corresponding to a maximally mixed logical output.

The effective physical error rate entering the outer code is modeled as Equation \ref{eq:10};
\begin{equation}
p_{\mathrm{phys}} = w_Z p_Z + w_P p_{\mathrm{dep}},
\label{eq:10}
\end{equation}
where $w_Z$ and $w_P$ weight residual dephasing and depolarizing contributions after GKP correction and stabilizer cycles.
In the numerical implementation, $w_Z = 0.3$ and $w_P = 0.1$, and $p_{\mathrm{phys}}$ is computed from the data-qubit noise channel only (ancilla and post-injection noise are not included in $p_{\mathrm{phys}}$).

\subsubsection{Fidelity renormalization}

Logical failures are conservatively assumed to produce a maximally mixed output with fidelity $1/2$. The final logical fidelity is therefore Equation \ref{eq:11}
\begin{equation}
F_{\mathrm{logical}}
= (1 - p_L) F_{\mathrm{inj}} + p_L \times \frac{1}{2}.
\label{eq:11}
\end{equation}

This approximation intentionally underestimates performance, ensuring that reported fidelities remain conservative.

\subsection{Sensitivity analysis}
To quantify robustness, finite-difference gradients of the logical fidelity with respect to loss and squeezing are computed (Equation~\ref{eq:12}):

\begin{subequations}
\begin{align}
\frac{\partial F_{\mathrm{logical}}}{\partial p_{\mathrm{base}}}
&\approx \frac{F(L+\Delta L) - F(L-\Delta L)}{2\Delta L}, \\
\frac{\partial F_{\mathrm{logical}}}{\partial s}
&\approx \frac{F(s+\Delta s) - F(s-\Delta s)}{2\Delta s}.
\end{align}
\label{eq:12}
\end{subequations}

These gradients are evaluated over the discrete parameter grid used in simulations and visualized as sensitivity heatmaps.

\subsection{Phase-boundary construction}
Finally, threshold-like operating regions are identified by imposing joint constraints in Equation~\ref{eq:13};
\begin{equation}
P_{\mathrm{succ}} \ge P^\ast, \qquad F_{\mathrm{logical}} \ge F^\ast,
\label{eq:13}
\end{equation}
with fixed target values $(P^\ast, F^\ast)$.

For each baseline loss and code distance, the minimum squeezing $s_{\min}$ satisfying both conditions is computed, yielding a phase boundary in Equation~\ref{eq:14};
\begin{equation}
s_{\min} = s_{\min}(p_{\mathrm{base}}, d).
\label{eq:14}
\end{equation}

This representation provides a compact design-space map linking hardware quality (squeezing and loss) to fault-tolerant performance requirements.

\subsection{Numerical implementation}
All simulations were performed using Monte Carlo sampling of density-matrix evolution under the effective noise channels described above. Each data point corresponds to $5\times10^3$ independent RUS trials, ensuring statistical convergence of success probabilities and fidelities. Random seeds were fixed for reproducibility. A single fixed input state $|\psi\rangle$ (drawn once at initialization and normalized) is used across the full sweep to ensure comparability between parameter settings.

\noindent
These methods enable a transparent and computationally efficient evaluation of fault-tolerant logical magic-state preparation in photonic GKP architectures, while retaining direct correspondence with experimentally tunable parameters.

\section{Results}
\label{sec:results}

\subsection{Architecture-level performance of GKP-based magic-state injection}
\label{subsec:performance}

Evaluation begins with the performance of a repeat-until-success (RUS) logical T-gate injection protocol implemented using GKP-encoded photonic qubits and protected by an outer surface code.The analysis is conducted at the architecture level, incorporating effective noise models for finite squeezing, photon loss, and logical depolarization, while avoiding explicit decoder or syndrome simulation.

\begin{figure*}[t]
    \centering
    \includegraphics[width=\linewidth]{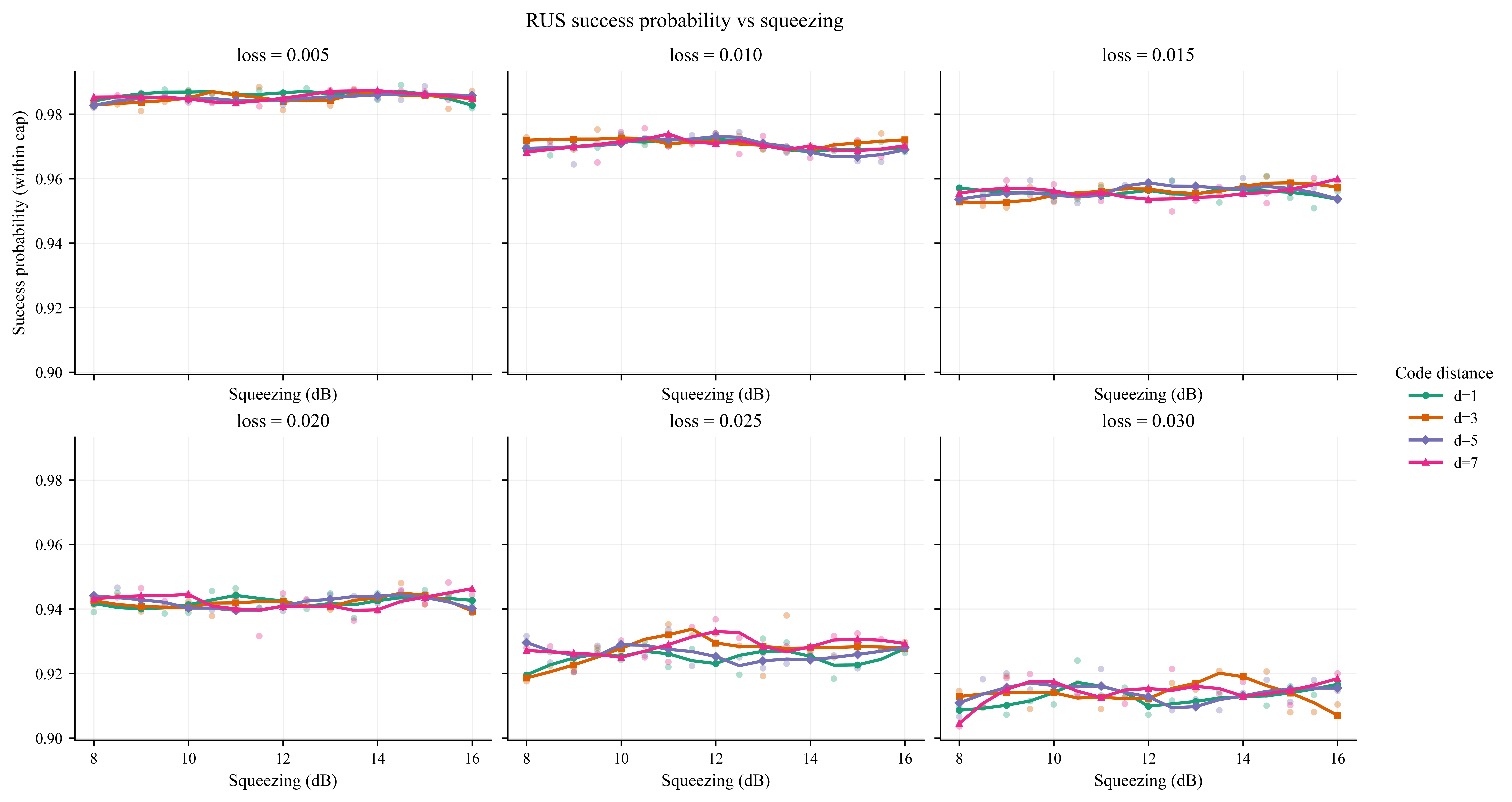}
    \caption{
    RUS success probability as a function of squeezing for baseline loss values $p_{\mathrm{base}}=0.005$--$0.03$ (panel titles).
    Each panel shows surface-code distances $d=1,3,5,7$; the success probability is high across the sweep and only weakly dependent on $d$.
    The vertical axis is zoomed to highlight small variations.
    }
    \label{fig:success_vs_squeezing}
\end{figure*}

Figure~\ref{fig:success_vs_squeezing} shows the RUS success probability as a function of the squeezing proxy for several baseline loss values and surface-code distances. Across all loss regimes considered, the success probability remains high ($\gtrsim 0.9$) and shows a weak upward trend with squeezing (within Monte Carlo fluctuations). Increasing the surface-code distance has only a weak effect on the success probability itself, indicating that the RUS dynamics are primarily governed by the injection circuit and heralded loss rather than by outer-code suppression.

\subsection{RUS overhead and repeat-until-success behavior}
\label{subsec:overhead}

\begin{figure*}[t]
    \centering
    \includegraphics[width=\linewidth]{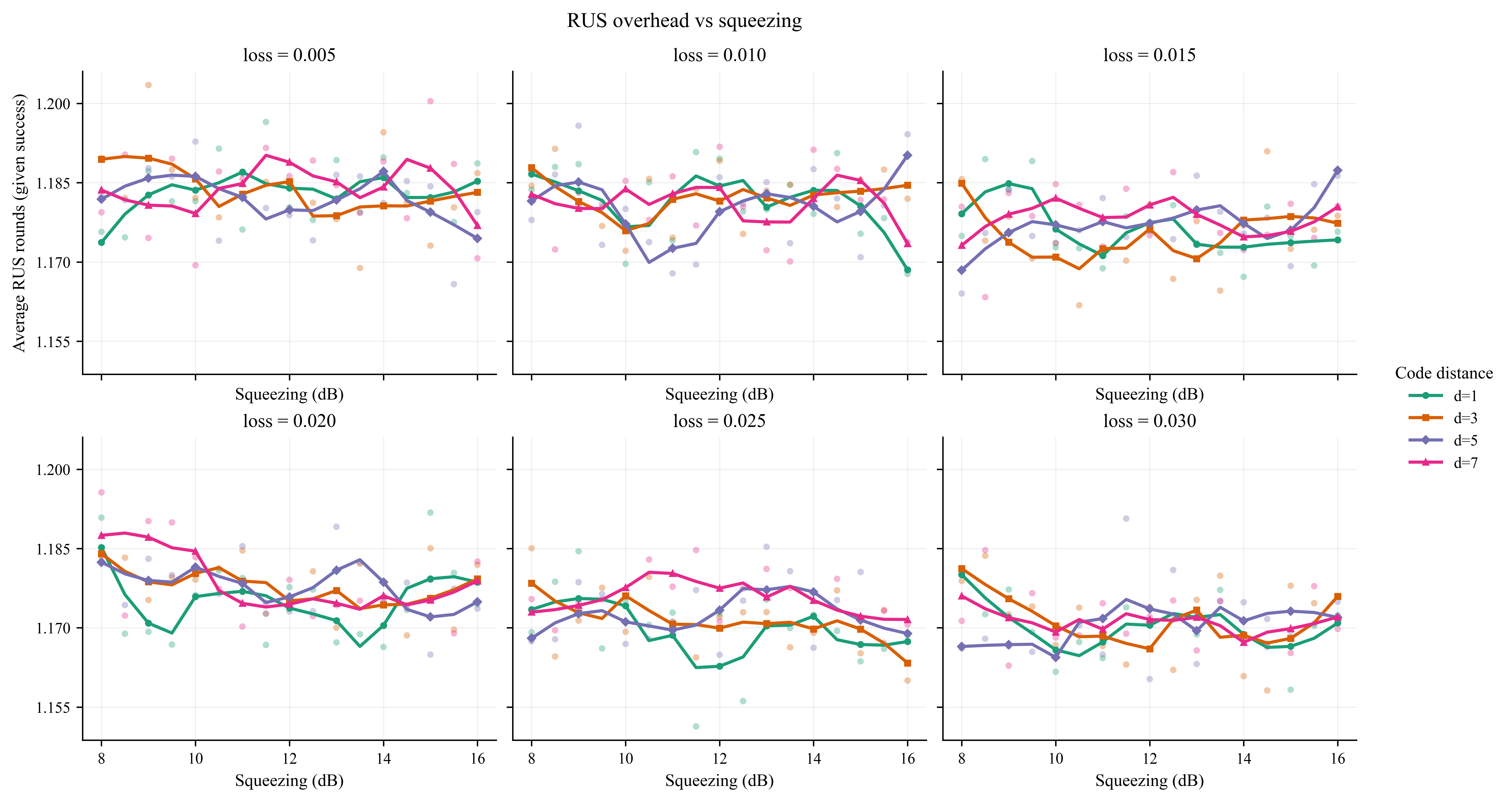}
    \caption{
    Average number of RUS rounds required for successful magic-state injection, conditioned on success, versus squeezing.
    Each panel corresponds to a baseline loss value, and curves indicate surface-code distance.
    The overhead remains near unity throughout, and the vertical axis is zoomed to reveal small differences.
    }
    \label{fig:overhead_vs_squeezing}
\end{figure*}

To quantify the operational overhead of the injection protocol, the average number of RUS rounds required to achieve a successful injection, conditioned on success, is examined. The results are shown in Fig.~\ref{fig:overhead_vs_squeezing}.

The average number of rounds remains close to unity across the entire parameter space, typically between $1.15$ and $1.21$. Notably, the overhead exhibits only a weak dependence on squeezing and loss, reflecting the fact that unsuccessful branches are efficiently heralded and restarted. Increasing the surface-code distance does not significantly increase the RUS overhead, demonstrating that logical protection does not introduce prohibitive latency in the injection process.

\subsection{Logical magic-state fidelity after outer-code protection}
\label{subsec:fidelity}

\begin{figure*}[t]
    \centering
    \includegraphics[width=\linewidth]{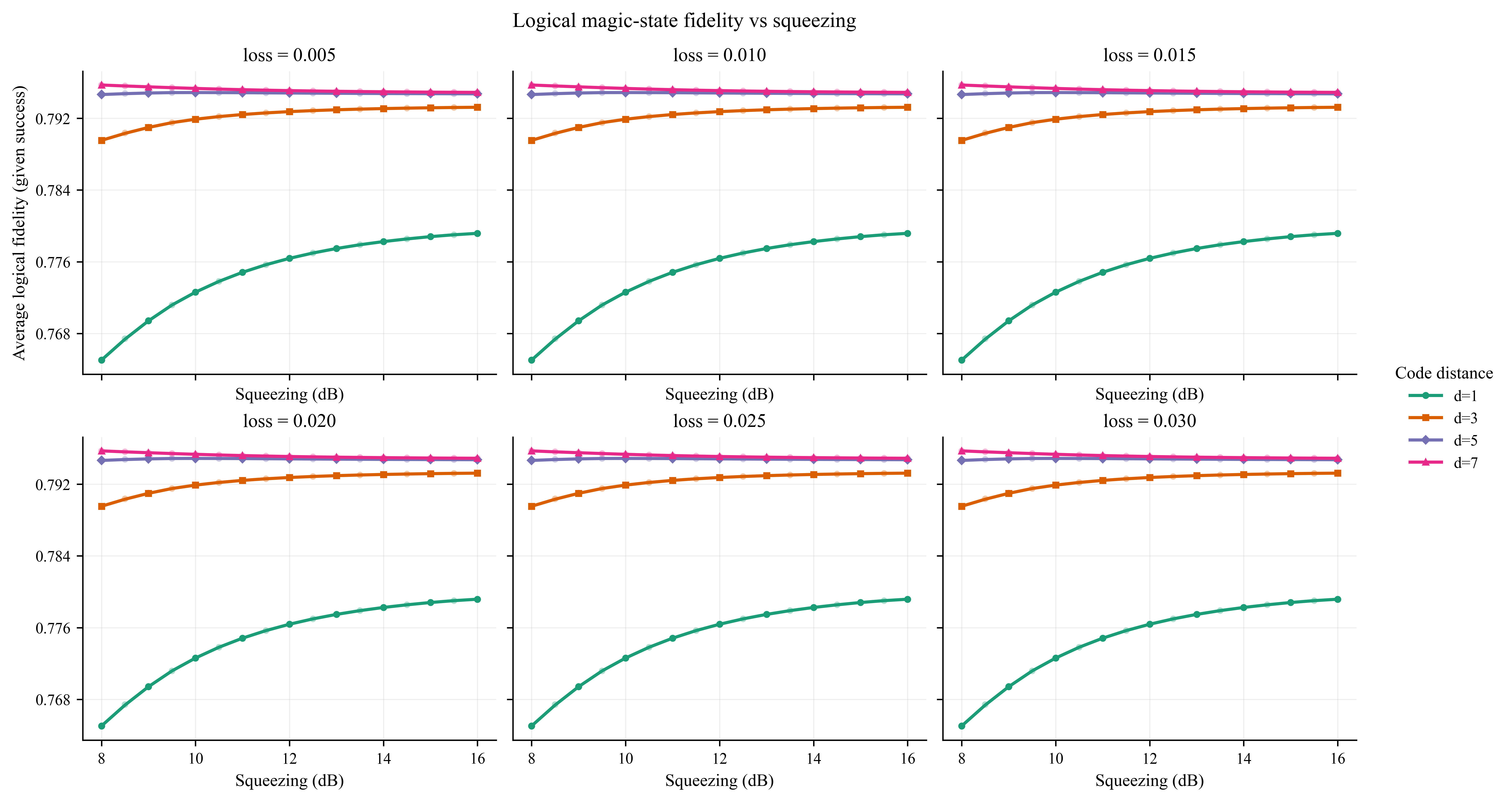}
    \caption{
    Average logical magic-state fidelity conditioned on successful injection as a function of squeezing.
    Panels sweep baseline loss and curves show surface-code distance; higher squeezing and larger $d$ increase logical fidelity, while moderate loss variations have a weak effect after outer-code protection.
    The vertical axis is zoomed to emphasize trends.
    }
    \label{fig:fidelity_vs_squeezing}
\end{figure*}

The central performance metric of the protocol is the logical fidelity of the injected magic state after outer-code protection. Figure~\ref{fig:fidelity_vs_squeezing} shows the average logical fidelity conditioned on successful injection, including the effect of surface-code error suppression.

For all loss values considered, the logical fidelity generally improves with squeezing, reflecting the reduction of effective GKP-induced phase noise. Increasing the surface-code distance systematically enhances the logical fidelity, with diminishing returns at higher distances, consistent with the expected scaling of surface-code logical error rates. Importantly, once outer-code protection is applied, the logical fidelity becomes largely insensitive to moderate variations in loss, indicating that photon loss is effectively converted into heralded erasure rather than uncorrectable logical errors.

\subsection{Sensitivity analysis}
\label{subsec:sensitivity}

To quantify the robustness of logical magic-state preparation and to identify dominant error sources, a sensitivity analysis of the logical fidelity with respect to both baseline photon loss and squeezing was performed. Finite-difference gradients of the logical fidelity were evaluated over the discrete architecture-level parameter grid explored in the simulations.

\begin{figure}[t]
    \centering
    \includegraphics[width=0.49\linewidth]{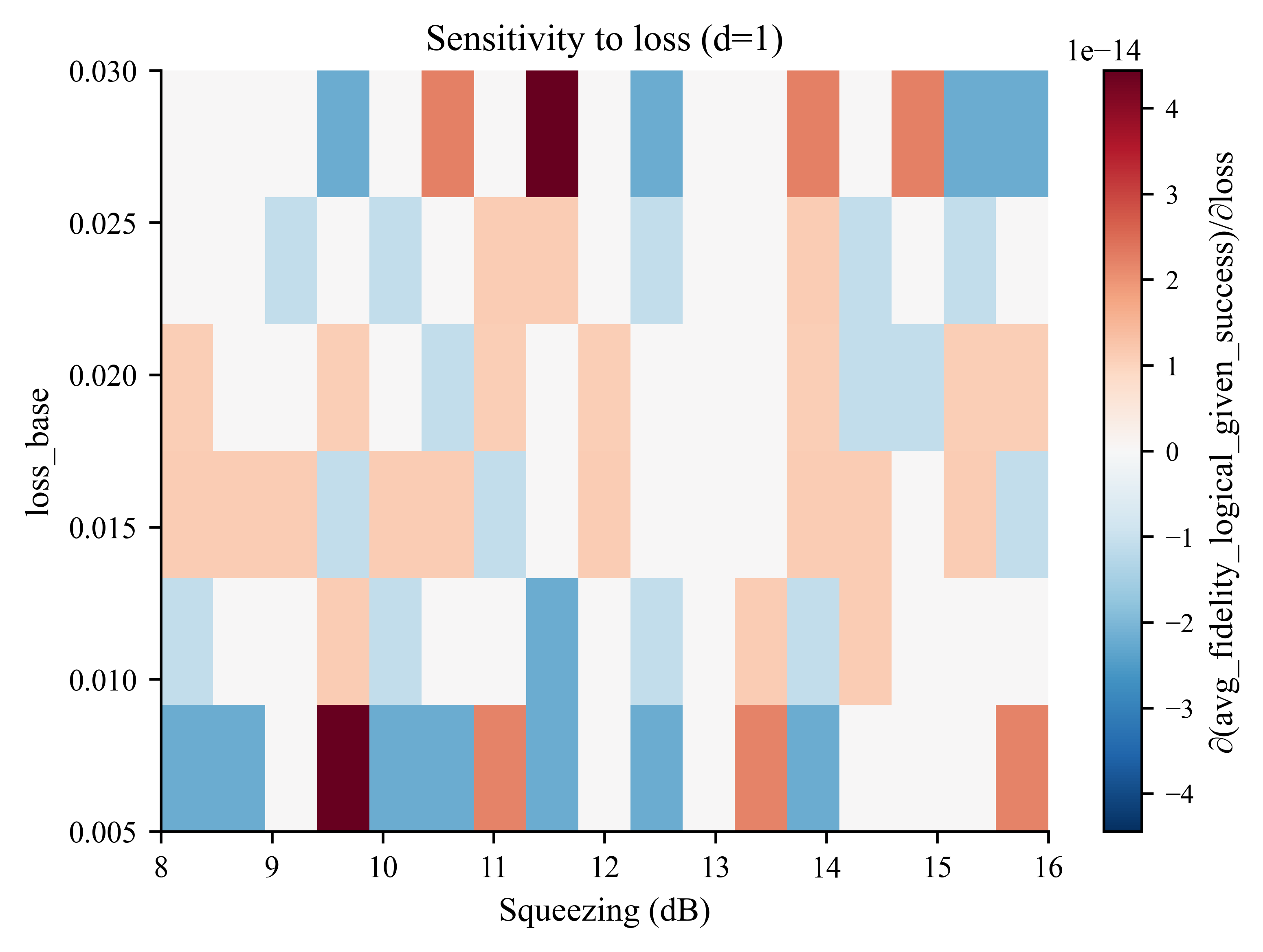}
    \includegraphics[width=0.49\linewidth]{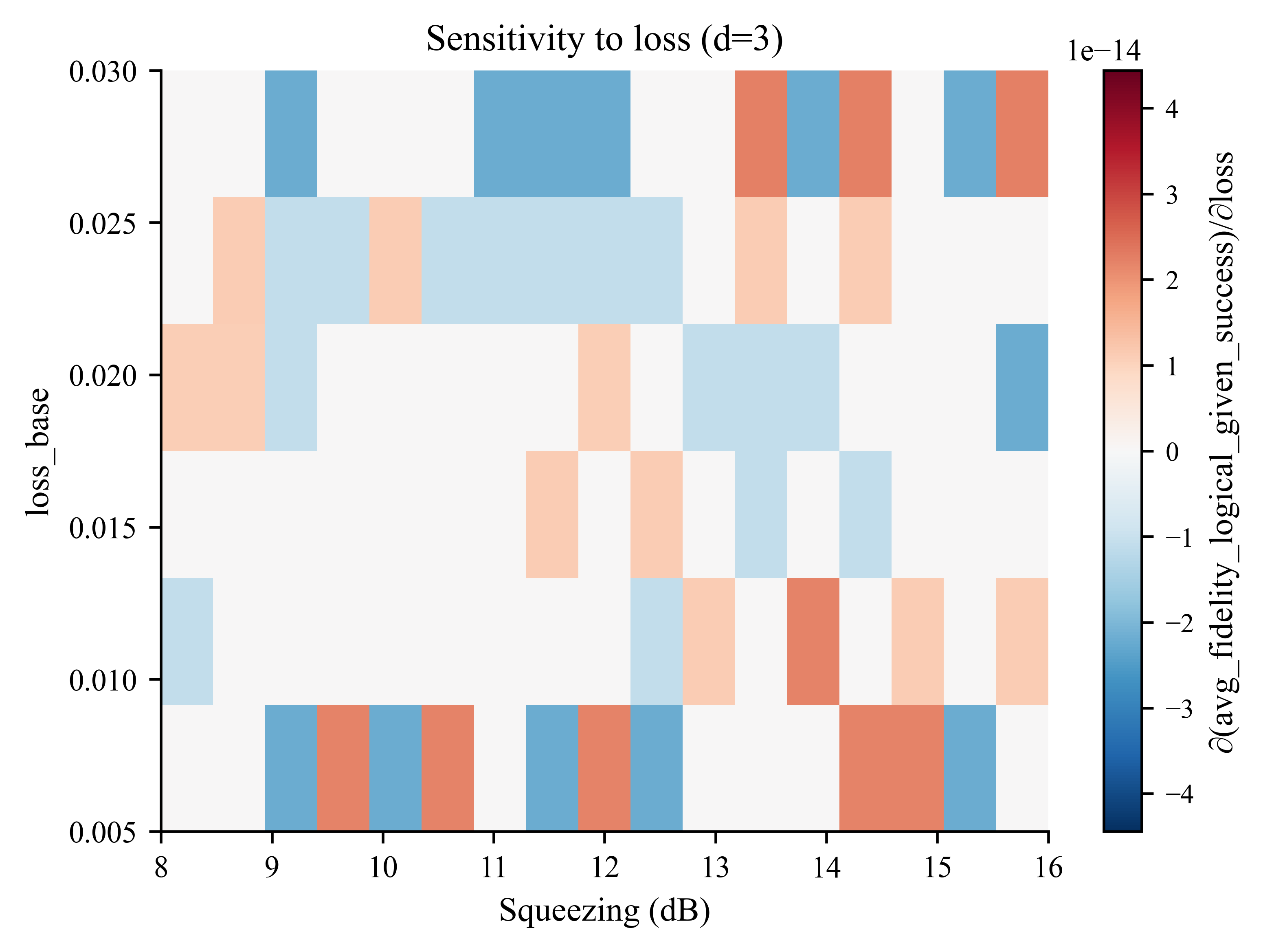}\\
    \includegraphics[width=0.49\linewidth]{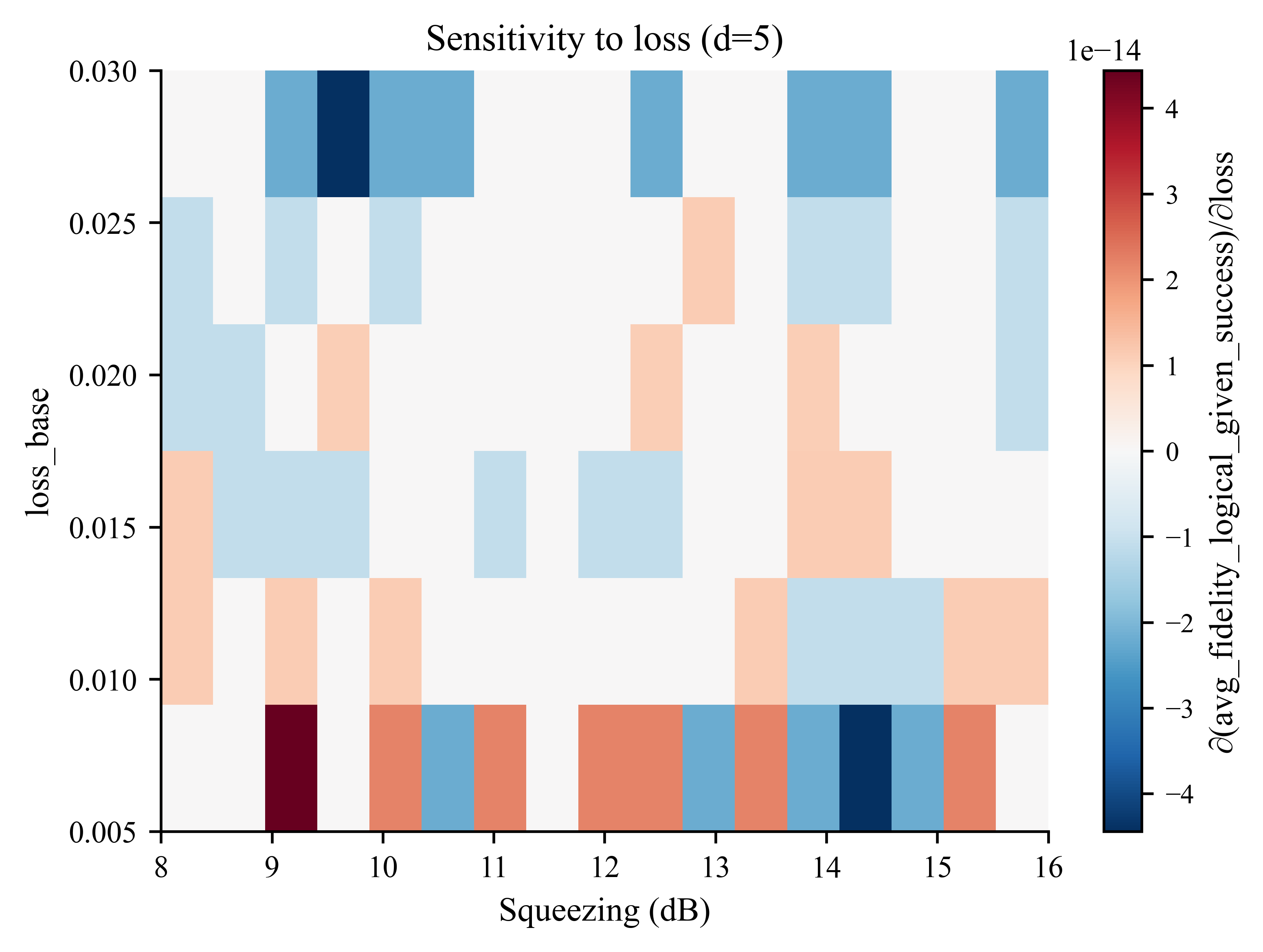}
    \includegraphics[width=0.49\linewidth]{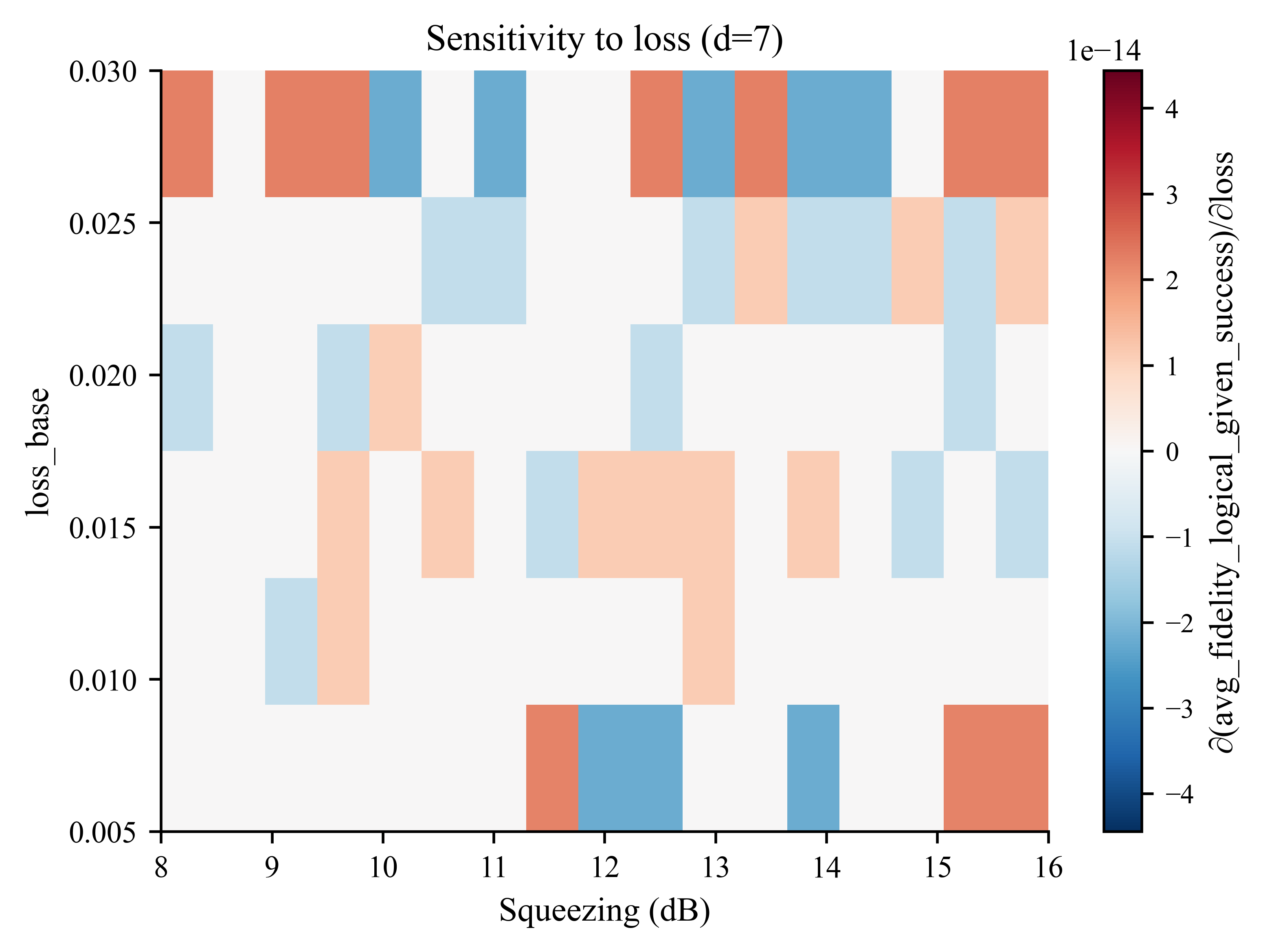}
    \caption{Sensitivity of logical fidelity to baseline photon loss, $\partial F_{\mathrm{log}}/\partial p_{\mathrm{base}}$, for surface-code distances $d=1,3,5,7$. Values remain near zero across the parameter space, indicating that heralded loss primarily affects success probability rather than the quality of successful logical states.}
    \label{fig:sensitivity_loss}
\end{figure}

\begin{figure}[t]
    \centering
    \includegraphics[width=0.49\linewidth]{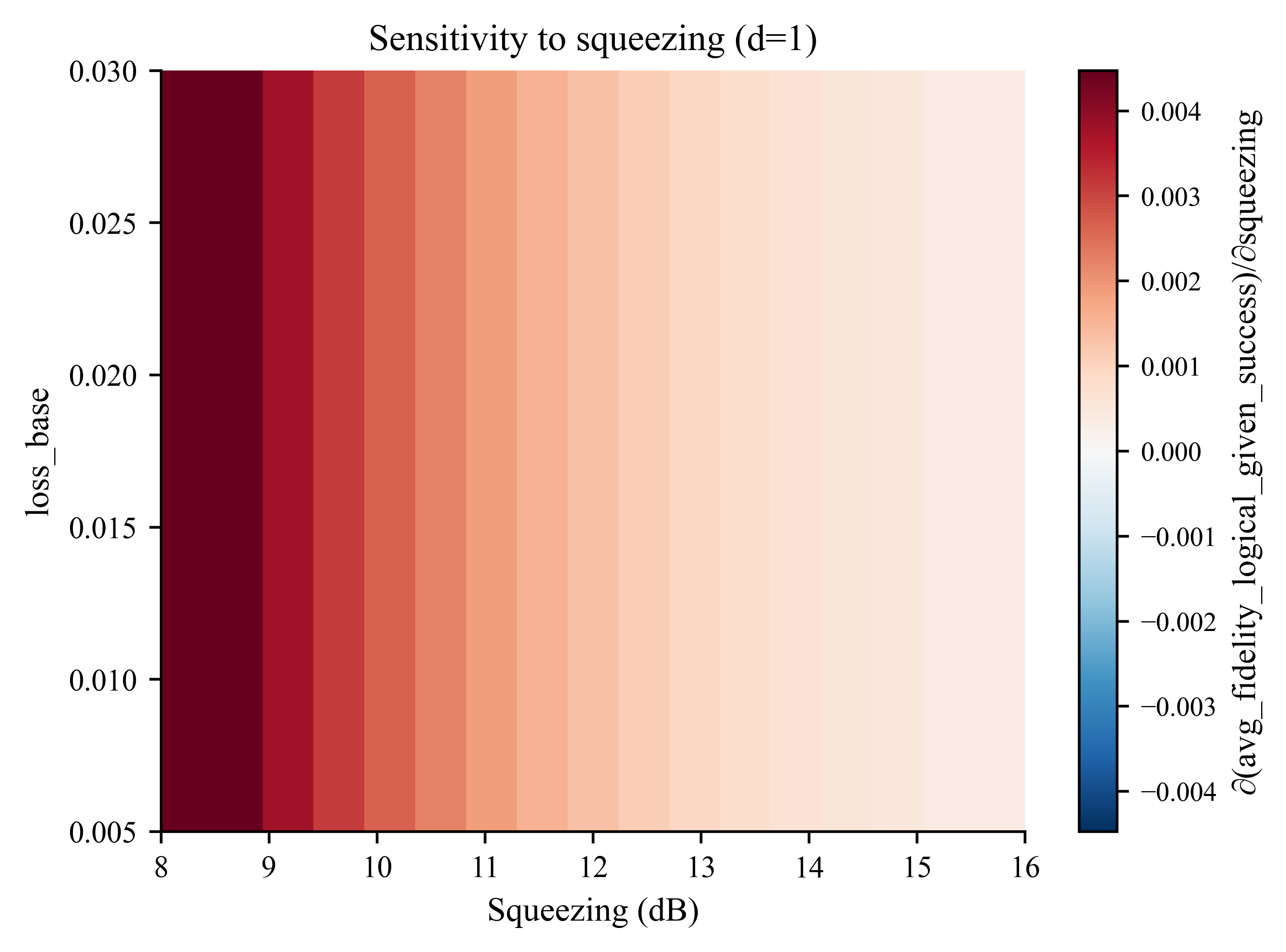}
    \includegraphics[width=0.49\linewidth]{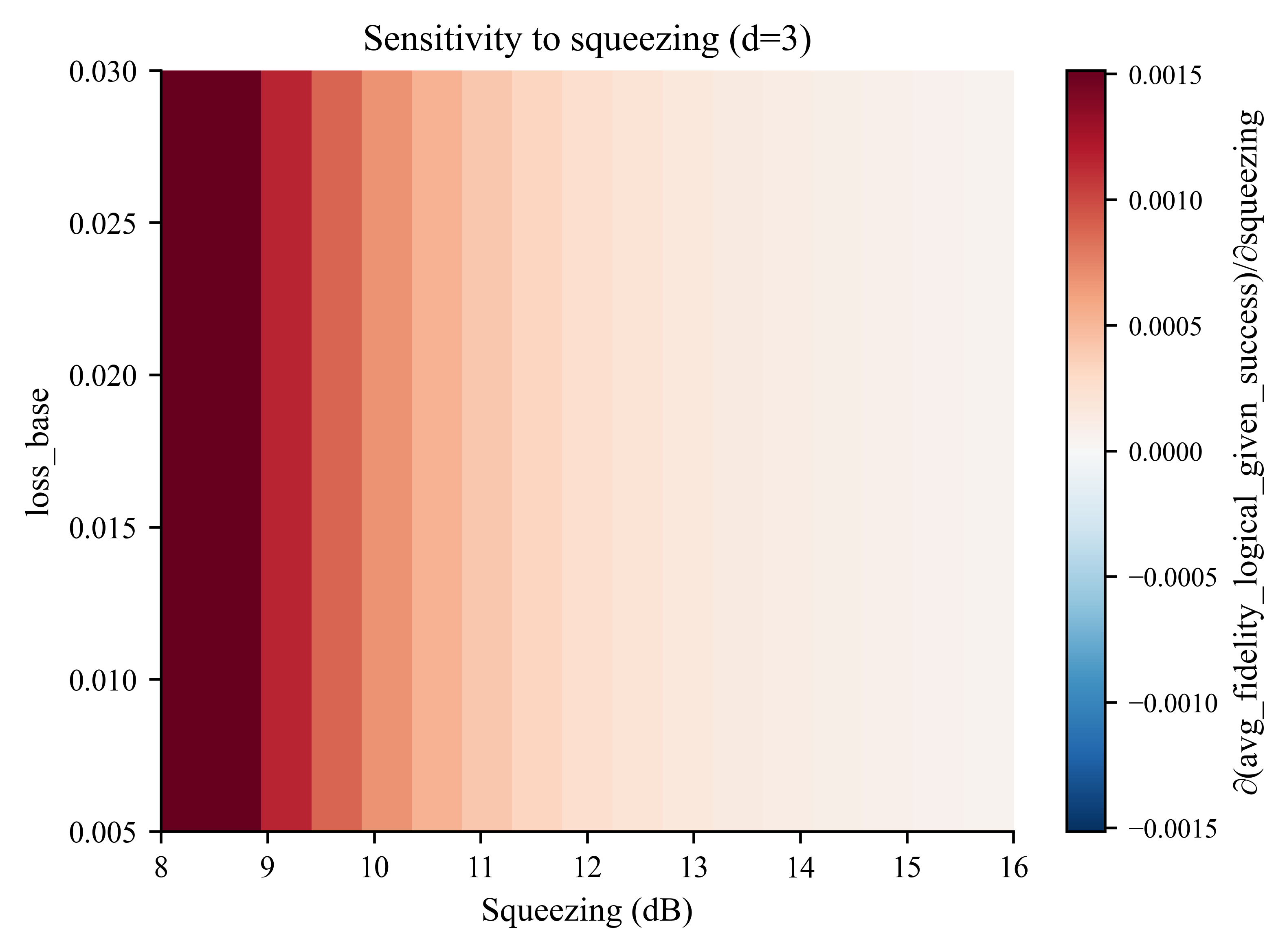}\\
    \includegraphics[width=0.49\linewidth]{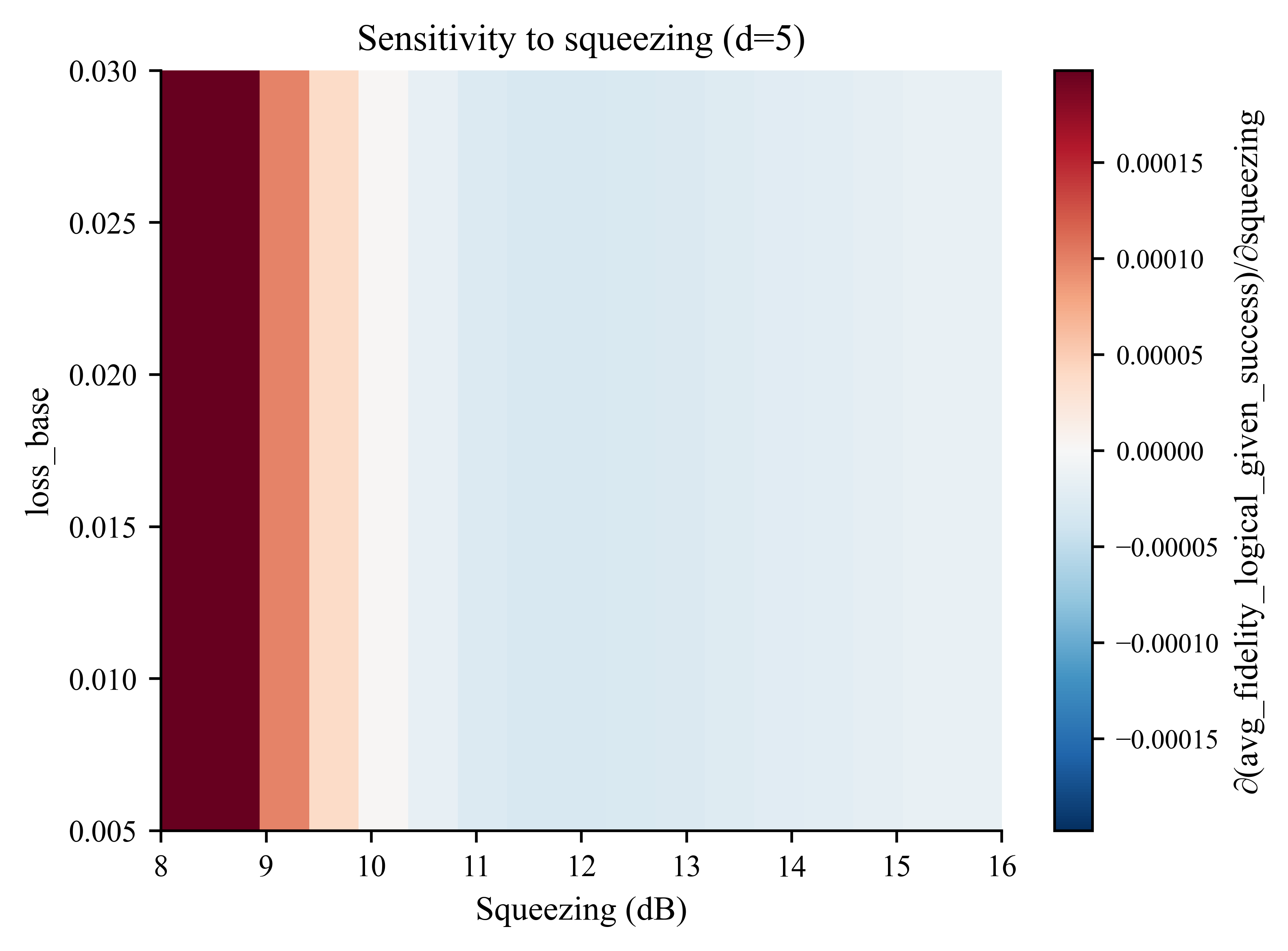}
    \includegraphics[width=0.49\linewidth]{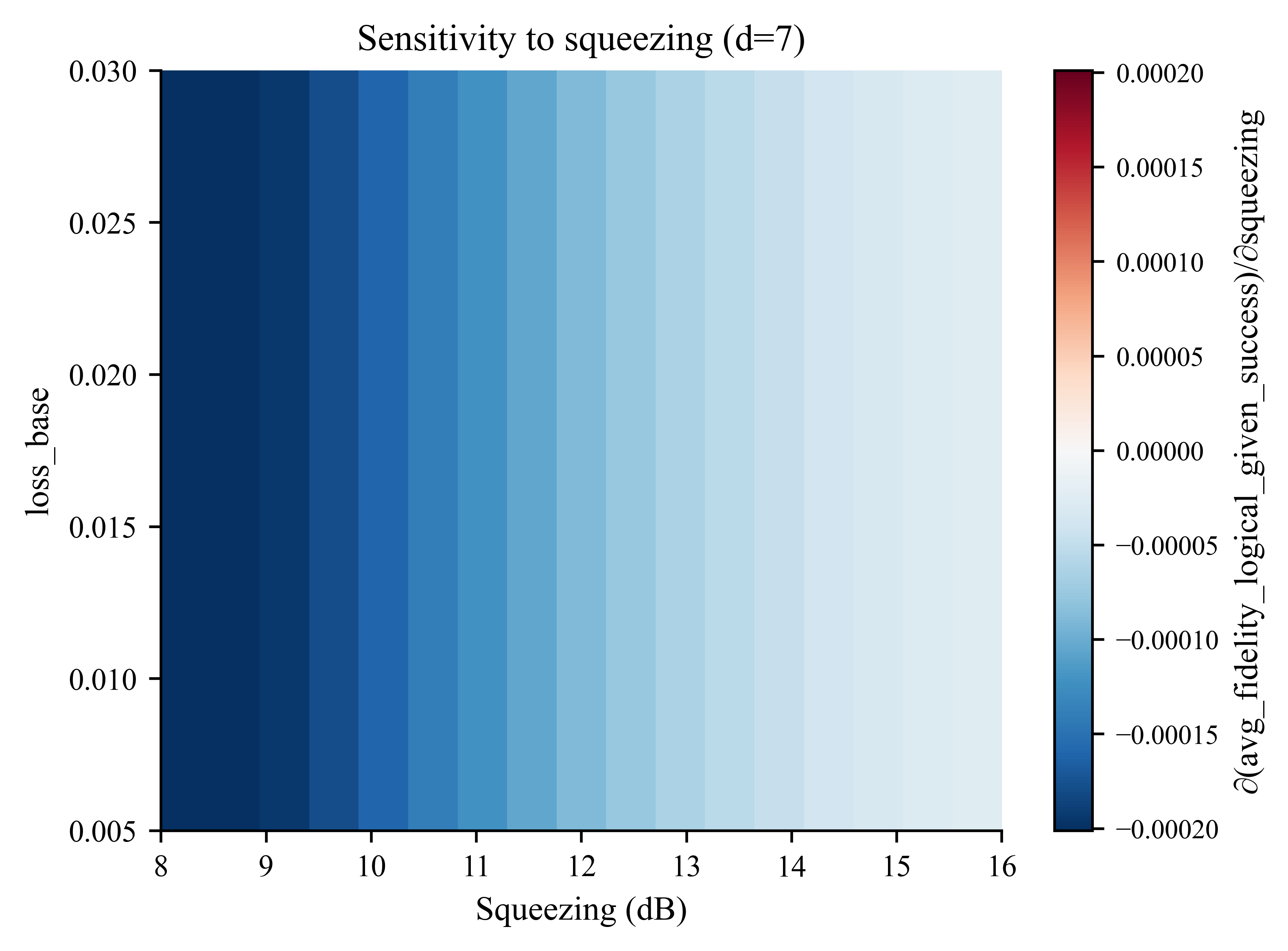}
    \caption{Sensitivity of logical fidelity to squeezing, $\partial F_{\mathrm{log}}/\partial s$, for surface-code distances $d=1,3,5,7$. Sensitivity is largest at low squeezing and diminishes with increasing squeezing and code distance, confirming finite-energy GKP noise as the dominant continuous error mechanism.}
    \label{fig:sensitivity_squeeze}
\end{figure}

Figures~\ref{fig:sensitivity_loss} and \ref{fig:sensitivity_squeeze} present sensitivity heatmaps for surface-code distances $d = 1, 3, 5,$ and $7$. Across all distances, the sensitivity of logical fidelity with respect to baseline loss remains close to zero throughout most of the parameter space, indicating strong robustness against moderate variations in photon loss. This behavior reflects the heralded nature of loss in photonic architectures, which primarily impacts success probability rather than the quality of successful logical states.

In contrast, the sensitivity with respect to squeezing is consistently nonzero and largest at low squeezing values. This confirms that finite-energy GKP noise constitutes the dominant continuous error mechanism affecting logical fidelity. Increasing the surface-code distance suppresses the overall magnitude of both sensitivities, further flattening the loss dependence while retaining a clear squeezing-driven performance gradient.

Results demonstrate that once combined with repeat-until-success injection and outer-code protection, photonic loss is effectively converted into a heralded failure channel, while squeezing quality directly governs the achievable logical-fidelity regime.

\subsection{Phase-boundary analysis}
\label{subsec:phase_boundary}
\begin{figure}[t]
    \centering
    \includegraphics[width=\linewidth]{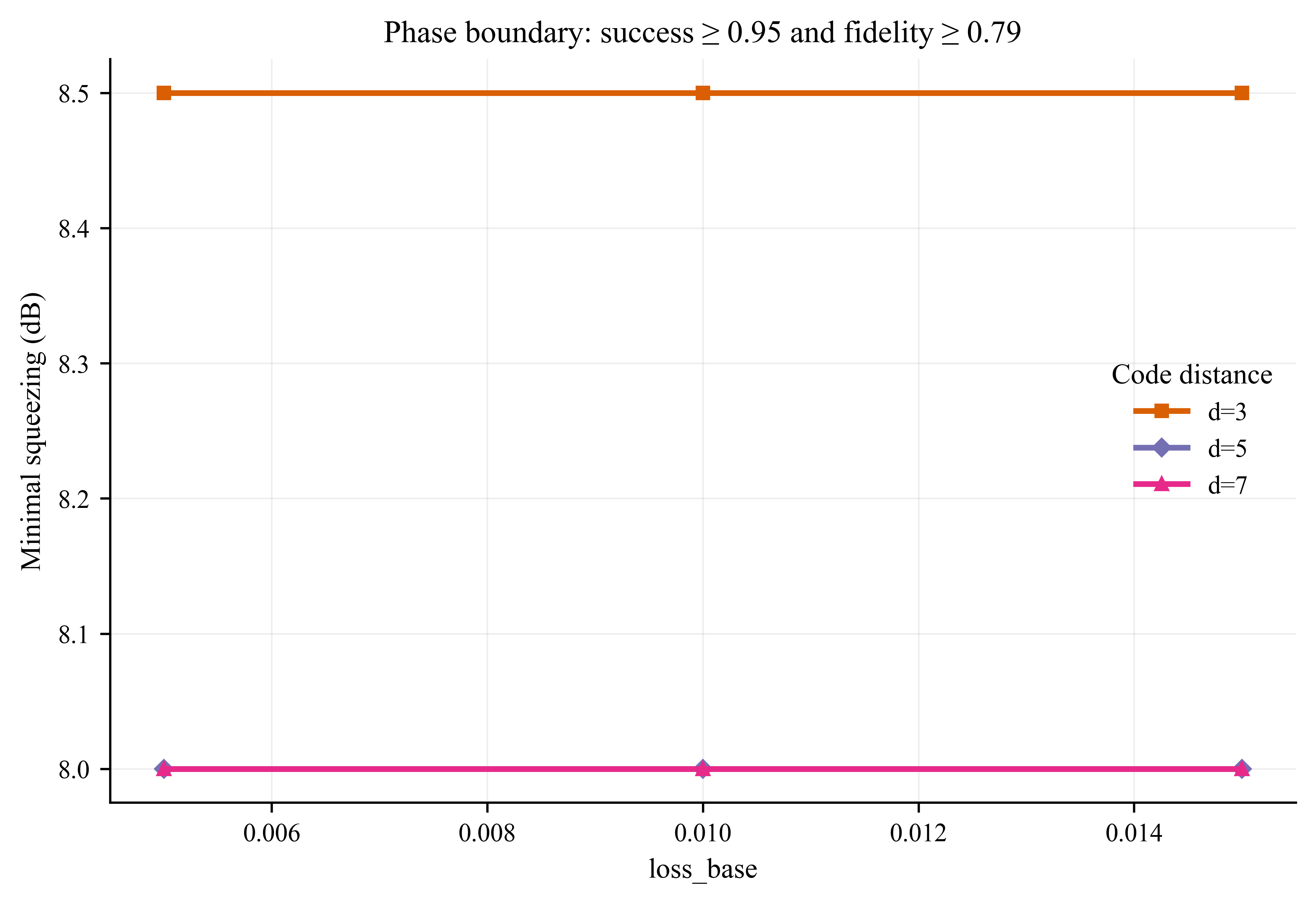}
    \caption{
    Phase boundary showing the minimum squeezing required to achieve both high RUS success probability ($\ge 0.95$) and high logical fidelity ($\ge 0.79$) as a function of baseline loss.
    Higher surface-code distances reduce the squeezing requirement, illustrating the trade-off between hardware quality and error-correction overhead; for $d=1$ no feasible points meet both thresholds in the simulated range.
    }
    \label{fig:phase_boundary}
\end{figure}

Finally, phase-boundary plots are constructed to identify the minimum squeezing required to simultaneously satisfy a target success probability and logical fidelity. Figure~\ref{fig:phase_boundary} shows the boundary defined by the conditions $P_{\mathrm{success}} \ge 0.95$, \qquad $F_{\mathrm{logical}} \ge 0.79$, for different surface-code distances.

The phase boundary reveals a clear trade-off between squeezing resources and outer-code distance. Increasing the code distance systematically shifts the boundary toward lower squeezing, demonstrating that stronger logical protection can partially compensate for limited squeezing resources; for $d=1$ no points satisfy both thresholds within the simulated range. This representation provides a compact and experimentally relevant design chart for photonic fault-tolerant architectures, enabling direct comparison of hardware capabilities and error-correction requirements.

\noindent
These results demonstrate that GKP-based photonic architectures can achieve robust, high-fidelity logical magic-state preparation using moderate squeezing and realistic loss levels, provided that heralded RUS protocols are combined with scalable outer-code protection.

\section{Discussions}

A major line of work on fault-tolerant photonic quantum computation has focused on the Gottesman–Kitaev–Preskill (GKP) encoding under finite squeezing, most notably in the seminal analysis by Menicucci et al. \cite{menicucci2014fault}, which established threshold theorems for continuous-variable cluster-state computation using idealized Gaussian operations. That study rigorously analyzed error propagation in measurement-based quantum computing and demonstrated that fault tolerance is achievable in principle with sufficiently high squeezing. However, the analysis primarily emphasized asymptotic thresholds and did not explicitly consider magic-state injection protocols, repeat-until-success dynamics, or the impact of heralded photon loss on gate-level overhead (Table~\ref{tab:lit_comparison}). In contrast, our work directly models logical $T$-gate magic-state preparation, explicitly incorporating finite squeezing, depolarizing noise, and erasure into a unified architecture-level framework. Rather than focusing on global thresholds, operational metrics—success probability, injection overhead, and logical fidelity—are quantified to provide performance measures relevant to near-term photonic architectures.

Table~\ref{tab:lit_comparison} summarizes how this work relates to representative prior studies on GKP fault tolerance, photonic architectures, and magic-state preparation.

\begin{table*}[t]
\caption{Comparison with representative literature on GKP fault tolerance, photonic architectures, and magic-state preparation.}
\label{tab:lit_comparison}
\centering
\scriptsize

\begin{tabular}{llllll}\toprule

Work & Platform / code & Primary focus & Non-Clifford handling & Loss treatment & Metrics \\
\midrule

Menicucci et al.\ (2014)~\cite{menicucci2014fault}
& \shortstack[l]{CV cluster states\\with GKP}
& \shortstack[l]{Fault-tolerance\\thresholds for MBQC}
& \shortstack[l]{Universal MBQC\\(no injection)}
& Not central
& \shortstack[l]{Thresholds vs\\squeezing} \\

Fukui et al.\ (2018)~\cite{fukui2018high}
& GKP with analog QEC
& Decoder-level thresholds
& Not explicit
& Not central
& Logical error rates \\

Noh \& Chamberland (2020)~\cite{noh2020fault}
& Surface--GKP code
& Bosonic FT QEC
& Not explicit
& Not central
& Logical error scaling \\

Nickerson et al.\ (2014)~\cite{nickerson2014freely}
& \shortstack[l]{Photonic modular\\architecture}
& Loss-tolerant networking
& Not explicit
& Erasure-aware
& \shortstack[l]{Loss thresholds, \\logical rates} \\

Bourassa et al.\ (2021)~\cite{bourassa2021blueprint}
& Photonic GKP architecture
& System-level blueprint
& Distillation at high level
& Loss-aware
& \shortstack[l]{Resource/\\overhead estimates} \\

Campbell et al.\ (2012)~\cite{campbell2012magic}
& Qubit codes (Reed--Muller)
& Magic-state distillation
& Distillation protocols
& Not central
& Distillation overheads \\

\textbf{This work}
& \shortstack[l]{GKP photonic +\\surface abstraction}
& \shortstack[l]{RUS-based injection\\architecture}
& Explicit RUS $T$ injection
& Heralded erasure
& \shortstack[l]{$P_{\mathrm{succ}}$, $\langle R\rangle_{\mathrm{succ}}$,\\ $F_{\mathrm{log}}$} \\ \bottomrule
\end{tabular}

\end{table*}

Recent studies by Fukui et al. \cite{fukui2018high} and related follow-up work introduced high-performance GKP error-correction schemes based on analog information, demonstrating that exploiting continuous measurement outcomes can substantially improve logical error suppression at moderate squeezing levels. These works provide important insights into decoder-level performance and show that finite squeezing noise can be mitigated more efficiently than previously thought. Related bosonic fault-tolerance analyses using surface--GKP codes, such as Noh and Chamberland \cite{noh2020fault}, report logical-error scaling but likewise do not model non-Clifford injection or loss in detail (Table~\ref{tab:lit_comparison}). Our approach complements these results by treating GKP protection as an effective logical noise map and explicitly studying how residual errors propagate through a non-Clifford injection protocol, thereby bridging the gap between decoder-centric analyses and system-level fault-tolerant gate synthesis.

Magic-state preparation has been studied extensively in qubit-based architectures, particularly through distillation protocols and measurement-based constructions \cite{campbell2009structure,bravyi2005universal,campbell2012magic,litinski2019game}. These studies emphasize theoretical optimality and overhead scaling under abstract noise models, and they do not explicitly incorporate photonic constraints such as finite squeezing or photon-loss-induced erasure (Table~\ref{tab:lit_comparison}). In contrast, our work focuses on injection rather than distillation and shows that, in photonic architectures, high logical fidelity can be achieved with modest overhead through RUS-style injection combined with outer-code protection within the simulated regime. This suggests an alternative pathway in which distillation-heavy pipelines could be reduced when injected-state quality is sufficiently high.

Surface-code-protected photonic architectures have been proposed by Nickerson, Li, and Benjamin \cite{nickerson2014freely}, who analyzed fault-tolerant quantum computation in modular and networked systems subject to photon loss. Their work demonstrated that heralded loss can be tolerated efficiently using erasure-aware decoding, significantly relaxing hardware requirements (Table~\ref{tab:lit_comparison}). However, their analysis focused primarily on Clifford operations and logical error rates, without explicitly examining non-Clifford gate synthesis or magic-state fidelity. Our results align with their conclusions regarding loss tolerance but extend them by showing that, once heralded loss is accounted for, logical magic-state fidelity is only weakly sensitive to moderate loss levels within the simulated range. Instead, finite squeezing emerges as the dominant factor limiting logical quality, a distinction that is made explicit through our sensitivity and phase-boundary analyses.

Finally, recent photonic fault-tolerance roadmaps, such as those developed by Rudolph \cite{rudolph2017optimistic} and subsequent architectural reviews \cite{bourassa2021blueprint,webster2022universal}, have emphasized the central role of squeezing as a hardware bottleneck while acknowledging uncertainty in how squeezing, loss, and code distance jointly constrain performance. These works provide broad qualitative guidance but stop short of quantitative design rules. Our phase-boundary diagrams address this gap by identifying the minimum squeezing required to simultaneously achieve high RUS success probability and high logical fidelity across different loss regimes and code distances. By mapping out these boundaries explicitly, our study helps translate qualitative architectural intuition into quantitative constraints, offering practical guidance for experimental photonic quantum computing platforms.

\section{Conclusion}

An architecture-level study of fault-tolerant logical magic-state preparation in GKP-encoded photonic quantum computing architectures has been presented. A repeat-until-success (RUS) logical $T$-gate injection protocol is combined with an abstracted outer surface-code layer to quantify how finite squeezing, photon-loss-induced erasure, and code distance jointly shape operational performance. This framework enables efficient exploration of a broad design space that would otherwise be prohibitively expensive to access using full continuous-variable wavefunction simulations or decoder-level surface-code modeling.

Within the simulated parameter ranges, logical magic-state injection is robust against moderate photon loss, with average RUS overheads remaining near unity even in relatively lossy settings. In contrast, finite squeezing is the dominant continuous error source affecting logical fidelity. Once protected by an outer code, logical magic-state fidelity shows only weak dependence on loss and diminishing gains with increasing code distance. Sensitivity analysis confirms this separation of roles: squeezing primarily governs logical quality, while loss predominantly impacts heralded success probabilities rather than post-selected fidelity.

Phase-boundary analysis reveals threshold-like boundaries in the simulated regime, identifying the minimum squeezing required to simultaneously achieve high success probability and high logical fidelity as functions of loss and code distance. These boundaries provide quantitative guidance for the co-design of photonic hardware and fault-tolerant logical architectures.

All simulations are performed using a custom density-matrix-based simulator implemented directly in Python and NumPy. No external quantum simulation frameworks, such as PennyLane, are employed. The framework is lightweight, transparent, and readily extensible, supporting rapid architectural exploration. These results clarify the relative importance of squeezing, loss, and code distance in photonic fault-tolerant quantum computing and provide a scalable modeling foundation that complements, rather than replaces, detailed device-level simulations.

\section*{Author Contributions}

D.D.K.W: conceptualization, methodology, validation and visualization, software, writing – original draft, review \& editing.

\section*{Acknowledgment(s)}

The author acknowledges the use of open-source scientific software in the development of this work. Numerical simulations and data analysis were carried out using \texttt{Python}, with core numerical routines implemented using \texttt{NumPy} and visualization performed using \texttt{Matplotlib}. Data handling and post-processing utilities relied on the standard \texttt{csv} and \texttt{dataclasses} libraries from the Python ecosystem. All simulations were executed on local computational resources.

Any opinions, findings, conclusions, or recommendations expressed in this research are those of the author and do not necessarily reflect the views of their respective affiliations.

\section*{Data \& Code Availability}
The data generated and analyzed during the present study are included within the manuscript. Supplementary codes developed for \texttt{LiDMaS} simulations are provided as supplementary material and accessible on \href{https://github.com/DennisWayo/lidmas-gkp}{GitHub} to ensure transparency and reproducibility.

\section*{Funding}
This research was not funded.

\section*{Disclosure statement}

No potential conflict of interest was reported by the author(s).

\bibliographystyle{apsrev4-2}
\bibliography{magic}

@article{menicucci2014fault,
  title={Fault-tolerant measurement-based quantum computing with continuous-variable cluster states},
  author={Menicucci, Nicolas C},
  journal={Physical review letters},
  volume={112},
  number={12},
  pages={120504},
  year={2014},
  publisher={APS}
}

@article{fukui2018high,
  title={High-threshold fault-tolerant quantum computation with analog quantum error correction},
  author={Fukui, Kosuke and Tomita, Akihisa and Okamoto, Atsushi and Fujii, Keisuke},
  journal={Physical review X},
  volume={8},
  number={2},
  pages={021054},
  year={2018},
  publisher={APS}
}

@inproceedings{campbell2009structure,
  title={On the structure of protocols for magic state distillation},
  author={Campbell, Earl T and Browne, Dan E},
  booktitle={Workshop on Quantum Computation, Communication, and Cryptography},
  pages={20--32},
  year={2009},
  organization={Springer}
}

@article{nickerson2014freely,
  title={Freely scalable quantum technologies using cells of 5-to-50 qubits with very lossy and noisy photonic links},
  author={Nickerson, Naomi H and Fitzsimons, Joseph F and Benjamin, Simon C},
  journal={Physical Review X},
  volume={4},
  number={4},
  pages={041041},
  year={2014},
  publisher={APS}
}

@article{rudolph2017optimistic,
  title={Why I am optimistic about the silicon-photonic route to quantum computing},
  author={Rudolph, Terry},
  journal={APL photonics},
  volume={2},
  number={3},
  year={2017},
  publisher={AIP Publishing}
}

@article{slussarenko2019photonic,
  title={Photonic quantum information processing: A concise review},
  author={Slussarenko, Sergei and Pryde, Geoff J},
  journal={Applied physics reviews},
  volume={6},
  number={4},
  year={2019},
  publisher={AIP Publishing}
}

@article{bourassa2021blueprint,
  title={Blueprint for a scalable photonic fault-tolerant quantum computer},
  author={Bourassa, J Eli and Alexander, Rafael N and Vasmer, Michael and Patil, Ashlesha and Tzitrin, Ilan and Matsuura, Takaya and Su, Daiqin and Baragiola, Ben Q and Guha, Saikat and Dauphinais, Guillaume and others},
  journal={Quantum},
  volume={5},
  pages={392},
  year={2021},
  publisher={Verein zur F{\"o}rderung des Open Access Publizierens in den Quantenwissenschaften}
}

@article{gottesman2001encoding,
  title={Encoding a qubit in an oscillator},
  author={Gottesman, Daniel and Kitaev, Alexei and Preskill, John},
  journal={Physical Review A},
  volume={64},
  number={1},
  pages={012310},
  year={2001},
  publisher={American Physical Society}
}

@article{noh2020fault,
  title={Fault-tolerant bosonic quantum error correction with the surface--Gottesman-Kitaev-Preskill code},
  author={Noh, Kyungjoo and Chamberland, Christopher},
  journal={Physical Review A},
  volume={101},
  number={1},
  pages={012316},
  year={2020},
  publisher={APS}
}

@article{bravyi2005universal,
  title={Universal quantum computation with ideal Clifford gates and noisy ancillas},
  author={Bravyi, Sergey and Kitaev, Alexei},
  journal={Physical Review A—Atomic, Molecular, and Optical Physics},
  volume={71},
  number={2},
  pages={022316},
  year={2005},
  publisher={APS}
}

@article{howard2014contextuality,
  title={Contextuality supplies the ‘magic’for quantum computation},
  author={Howard, Mark and Wallman, Joel and Veitch, Victor and Emerson, Joseph},
  journal={Nature},
  volume={510},
  number={7505},
  pages={351--355},
  year={2014},
  publisher={Nature Publishing Group UK London}
}

@article{campbell2012magic,
  title={Magic-state distillation in all prime dimensions using quantum reed-muller codes},
  author={Campbell, Earl T and Anwar, Hussain and Browne, Dan E},
  journal={Physical Review X},
  volume={2},
  number={4},
  pages={041021},
  year={2012},
  publisher={APS}
}

@article{litinski2019game,
  title={A game of surface codes: Large-scale quantum computing with lattice surgery},
  author={Litinski, Daniel},
  journal={Quantum},
  volume={3},
  pages={128},
  year={2019},
  publisher={Verein zur F{\"o}rderung des Open Access Publizierens in den Quantenwissenschaften}
}

@article{ralph2005loss,
  title={Loss-tolerant optical qubits},
  author={Ralph, Timothy C. and Hayes, Alex J. and Gilchrist, Alexei},
  journal={Physical Review Letters},
  volume={95},
  number={10},
  pages={100501},
  year={2005},
  publisher={American Physical Society}
}

@article{fowler2012surface,
  title={Surface codes: Towards practical large-scale quantum computation},
  author={Fowler, Austin G. and Mariantoni, Matteo and Martinis, John M. and Cleland, Andrew N.},
  journal={Physical Review A},
  volume={86},
  number={3},
  pages={032324},
  year={2012},
  publisher={American Physical Society}
}

@article{webster2022universal,
  title={Universal fault-tolerant quantum computing with stabilizer codes},
  author={Webster, Paul and Vasmer, Michael and Scruby, Thomas R and Bartlett, Stephen D},
  journal={Physical Review Research},
  volume={4},
  number={1},
  pages={013092},
  year={2022},
  publisher={APS}
}

@article{divincenzo2009fault,
  title={Fault-tolerant architectures for superconducting qubits},
  author={DiVincenzo, David P},
  journal={Physica Scripta},
  volume={2009},
  number={T137},
  pages={014020},
  year={2009},
  publisher={IOP Publishing}
}

@article{bergholm2018pennylane,
  title={Pennylane: Automatic differentiation of hybrid quantum-classical computations},
  author={Bergholm, Ville and Izaac, Josh and Schuld, Maria and Gogolin, Christian and Ahmed, Shahnawaz and Ajith, Vishnu and Alam, M Sohaib and Alonso-Linaje, Guillermo and AkashNarayanan, B and Asadi, Ali and others},
  journal={arXiv preprint arXiv:1811.04968},
  year={2018}
}

@article{hastrup2021improved,
  title={Improved readout of qubit-coupled Gottesman--Kitaev--Preskill states},
  author={Hastrup, Jacob and Andersen, Ulrik Lund},
  journal={Quantum Science and Technology},
  volume={6},
  number={3},
  pages={035016},
  year={2021},
  publisher={IOP Publishing}
}

@article{paetznick2013repeat,
  title={Repeat-until-success: Non-deterministic decomposition of single-qubit unitaries},
  author={Paetznick, Adam and Svore, Krysta M.},
  journal={Quantum Information \& Computation},
  volume={14},
  number={15--16},
  pages={1277--1301},
  year={2014}
}

@book{gottesman1997stabilizer,
  title={Stabilizer codes and quantum error correction},
  author={Gottesman, Daniel},
  year={1997},
  publisher={California Institute of Technology}
}

\end{document}